\documentclass[prb,preprint,superscriptaddress,showpacs,amsmath,amssymb,nofootinbib,longbibliography]{revtex4-1}

\usepackage{multirow}
\usepackage{graphicx}
\usepackage{hyperref}
\usepackage{amsmath}
\usepackage{amssymb}
\usepackage{dcolumn}
\usepackage{url}
\usepackage{pdfpages}

\newcommand*{\citen}[1]{%
  \begingroup
    \romannumeral-`\x 
    \setcitestyle{numbers}%
    \cite{#1}%
  \endgroup   
}

\begin{document}

\title{Attending to experimental physics practices and lifelong learning skills in an introductory laboratory course}
\author{Punit R. Gandhi}
\email{punit\_gandhi@berkeley.edu}
\affiliation{Department of Physics, University of California,
Berkeley, Berkeley, CA 94720, USA}

\author{Jesse A. Livezey}
\affiliation{Department of Physics, University of California, Berkeley, Berkeley, CA 94720, USA}
\author{Anna M. Zaniewski}
\affiliation{Department of Physics, Arizona State University, Tempe, AZ 85287, USA}
\author{Daniel L. Reinholz}
\affiliation{Center for STEM Learning, University of Colorado Boulder, Boulder, CO 80309, USA}
\author{Dimitri R. Dounas-Frazer}
\affiliation{Department of Physics, University of Colorado Boulder, Boulder, CO 80309, USA}

\date{\today}

\begin{abstract}
We have designed an introductory laboratory course that engaged first-year undergraduate students in two complementary types of iteration: (1) \emph{iterative improvement of experiments} through cycles of modeling systems, designing experiments, analyzing data, and refining models and designs; and (2) \emph{iterative improvement of self} through cycles of reflecting on progress, soliciting feedback, and implementing changes to study habits and habits of mind. The course consisted of three major activities: a thermal expansion activity, which spanned the first half of the semester; final research projects, which spanned the second half of the semester; and guided student reflections, which took place throughout the duration of the course. We describe our curricular designs and report examples of student work that demonstrate students' iterative improvements in multiple contexts.
\end{abstract}

\maketitle

\section{Introduction}\label{sec:intro}

The design and study of undergraduate laboratory courses is an area of increasing interest in the physics education community, both at introductory\cite{Brewe2008,Etkina2007} and upper-division\cite{Zwickl2014,Zwickl2013AJP} levels. Recently, the American Association of Physics Teachers released recommendations for learning outcomes of undergraduate laboratory courses, which included the ability to model systems, design experiments, and analyze data.\cite{AAPT2015} These recommendations are in line with the cognitive tasks required for table-top atomic physics experiments\cite{Wieman} as well as the scientific and engineering practices outlined in the \emph{Next Generation Science Standards} for K-12 education.\cite{NGSS2013} More broadly, there is growing recognition that to support student success, physics education must focus on ``lifelong learning skills" like organization,\cite{Reid2008} collaboration,\cite{Dounas-Frazer2015} emotional regulation,\cite{Jaber2014} self-learning,\cite{NRC2013} and other social and interpersonal aspects of learning.\cite{NCES2011,Clark2005}

We designed an introductory laboratory course that engaged first-year undergraduate students in two types of iteration: (1) \emph{iterative improvement of experiments} through cycles of modeling systems, designing experiments, analyzing data, and refining models and designs; and (2) \emph{iterative improvement of self} through cycles of reflecting on progress, soliciting feedback, and implementing changes to study habits and habits of mind. This course differs from typical laboratory courses in two ways. First, it is an \emph{introductory} course that focuses on the iterative process of developing, testing, and refining models and apparatus---practices which typically feature in upper-division courses\cite{Zwickl2015,Dounas-Frazer2015arXiv} rather than at the introductory level\cite{Wieman} (though notable exceptions exist; \emph{e.g.} refinement is a major feature of introductory ISLE labs).\cite{Etkina2008,Etkina2009} Second, it incorporates a guided reflection activity\cite{Dounas-Frazer2015} that explicitly recognizes and attends to social and interpersonal aspects of learning. This explicit attention is in contrast to most courses, introductory or otherwise, which rely on a ``hidden curriculum'' that students must master in order to be successful.\cite{Jackson1968}

In this paper, we describe our curriculum in detail using examples of student work to help clarify the nature of the course. Evaluation of the impacts of the course is beyond the scope of the present work. Our discussion is organized as follows: programmatic context for the course is provided in Sec.~\ref{sec:context}; the design and implementation of activities related to our two course goals are described in Sec.~\ref{sec:design}; and Sec.~\ref{sec:concl} provides a summary of our work. Throughout Sec.~\ref{sec:design}, we provide examples of student work to illustrate how {Intro to Measurement} engaged students in iterative improvement of experiment and self.

\section{Programmatic Context}
\label{sec:context}


We describe a course, \emph{Intro to Measurement}, that was designed\footnote{One of the authors (DRDF) co-designed and co-taught the initial iteration of {Intro to Measurement} in the spring semester of 2012, and two of the authors (PRG and JAL) refined and co-taught the course the following spring. We report on the implementation and pilot study of the second iteration of the course.} as part of The Compass Project at the University of California, Berkeley (hereafter, ``Compass"). Established in 2006, Compass is a student-led initiative that aims to increase retention of students from underrepresented groups in the physical sciences by fostering a supportive community of graduate and undergraduate students.\cite{Albanna2013}  Compass offers three major programs to first- and second-year undergraduate students: a residential summer program for incoming students (cf. Ref~[\citen{Dounas-Frazer2013TPT}]); academic-year coursework that engages students in model-based reasoning and self-reflection (elaborated herein); and mentorship that starts in students' first year and may continue throughout their undergraduate experience. Students who complete the summer program are strongly encouraged to participate in subsequent coursework and mentorship.

Undergraduate students in Compass comprise a group that is more diverse than the Berkeley Physics Department as a whole. For example, of the 134 students who completed the Compass Summer Program from 2007--15, 43\% were women, 40\% were from underrepresented minority groups, and 31\% were first-generation college students. In contrast, in 2012 only 21\% of Berkeley physics majors were women and 7\% were underrepresented minorities, consistent with national trends.\cite{AIP2012}

We report student work from the second implementation of {Intro to Measurement}, during which 10 students were enrolled in the course. Eight students consented to use of their coursework for publication purposes.\footnote{Consent was obtained pursuant to protocol \#2012-11-4836, approved by the UC Berkeley Committee for Protection of Human Subjects.} Enrollment was open to students who completed the Summer Program as well as first-year students from the Physics Department more generally. As a result, representation of women and underrepresented minority students in this small population was more similar to the demographics of the Physics Department than those of Compass: of 8 research participants, 2 were women, 1 was a member of an underrepresented minority group, and 2 abstained from reporting demographic information. In addition, of the 6 students who provided demographic information, 3 reported having a disability, 1 identified as a first-generation college student, and 1 identified as a gender or sexual minority.

The structure of both organizational and educational aspects of Compass is informed by four community values: student ownership, collaboration, supporting students as ``whole people," and improvement through iteration.\cite{Albanna2013}  While each of these values features in {Intro to Measurement}, herein we focus on the ways in which our course engaged students in iterative improvement.

\section{Curriculum Design}
\label{sec:design}

{Intro to Measurement} met for two hours once weekly for a 14-week semester (28 hours total) with 10 students and two graduate student co-instructors. The course was designed for students to spend an additional three hours per week working on assignments outside of class (42 hours total). Two major objectives of the course were for students to engage in: 

(1) iterative improvement of experiments, through cycles of designing and refining systems and system models; and (2) iterative improvement of self, through cycles of reflection, feedback, and growth.

Our curricular design was influenced by \emph{Complex Instruction},\cite{Cohen1997} a set of strategies which promote equity in heterogeneous classrooms, and \emph{Constructionism},\cite{Papert1991} a philosophy which emphasizes learning through the construction of public artifacts that are shared with a community of peers. Along these lines, students worked in groups with well-defined roles, experimental tasks were challenging and open-ended, and the course culminated in a final research project and poster session---features that can also be found in other non-traditional introductory lab courses.\cite{Planinsic2007} Further, the course was designed to promote self-learning\cite{NRC2013} and provide students with social and interpersonal support.\cite{NCES2011} To this end, students were regularly required to submit written reflections on their growth as learners and they received individualized instructor feedback on each reflection.

In this section, we describe the curriculum in three parts (Fig.~\ref{fig:1}): a thermal expansion activity, which spanned the first half of the semester; final research projects, which spanned the second half of the semester; and guided student reflections, which took place throughout the first 11 weeks of the course. In each subsection, we provide examples of student work, such as the particular revisions students made to the thermal expansion apparatus and samples of student reflections.

\subsection{Thermal expansion activity}

During the first half of the semester, students performed experiments with a thermal expansion apparatus that has been described in detail elsewhere.\cite{Dounas-Frazer2013AJP} The purpose of these experiments was to provide students with opportunities to engage in the iterative process of designing, conducting, and refining experiments, in alignment with our first course objective.

The experimental apparatus consisted of a taut horizontal wire from which a hanging weight was suspended (Fig.~\ref{fig:2}). The wire expanded when heated, causing the hanging weight to drop lower to the ground. By measuring the temperature of the wire and change in height of the mass, one could use this apparatus to determine the wire's coefficient of thermal expansion. However, the wire's elasticity, mass, and kinks contribute to important systematic effects that may need to be taken into account depending on the desired accuracy and precision with which the thermal expansion coefficient is to be measured.\cite{Dounas-Frazer2013AJP}  This experimental apparatus is appropriate for an introductory lab course because the underlying principles and concepts (gravity, tension, elastic stretching, thermal expansion, and Newton's Second Law) are accessible to students with an introductory physics background. The apparatus was the focus of several complementary activities in which students engaged during this first phase of the course: exploration of the apparatus and related concepts; development, testing, and refinement of  models,  apparatus, and data-taking procedures; discussions about the nature of research; and inquiry-based data analysis tutorials.

During the initial exploration activity (Class~1), students rotated through five hands-on research stations. At these stations, students:
\begin{enumerate}
\item used a metal ring, metal ball, liquid nitrogen, and blow torch to gain experience with the phenomenon of thermal expansion;
\item used a rubberband, heating lamp, and computer-aided data acquisition to gain experience with  the contraction of materials upon heating;
\item used thermocouples to measure resistive heating by taking temperature measurements of a wire that was connected to a voltage source;
\item measured changes in the stretched length of a vertically hung rubberband when loads of different masses were attached to the bottom end; and
\item observed and documented the shapes of hanging chains, wires, and strings when loads of different masses (including zero mass, i.e., no load) were attached to the midpoint.
\end{enumerate}
Each station was accompanied by a short description of the phenomenon of interest as well as instructions for relevant equipment. These investigations provided shared experiences students could draw upon to be successful later in the course, reducing disparities in prior knowledge.  For example, the fifth station provided a foothold for future class discussions about whether and when the mass of the load in the thermal expansion experiment was ``heavy enough" that the wire's mass could be neglected.

In Class~2, students were given the thermal expansion apparatus (Fig.~\ref{fig:2}) and tasked with designing their own experimental procedure to measure the thermal expansion coefficient of the wire. Students used their experimental procedures to collect an initial round of data the following week (Class~3). In Class~4, they developed an initial theoretical model to analyze their data. The model, which was algebraic in nature and neglected the mass and elasticity of the wire\cite{Dounas-Frazer2013AJP}, was developed through a combination of small- and large-group discussions. Throughout this process, instructors engaged students in data analysis tutorials both in and out of class. Tutorials focused on basic statistical concepts (like averages, variances, and uncertainty propagation) as well as on the use of spreadsheet software to automate analysis of data.


Prior to Class~5, students' initial theoretical model neglected the wire's mass and elasticity. In Classes~5--6, students used their previously collected data to inform refinements to their models and experimental procedures. Refinements were discussed and agreed upon in small- and large-group discussions. Prior to discussions, instructors seeded ideas about which assumptions of the model to question by making anecdotal observations during data collection, such as noting that the wire hangs in a triangle even at room temperature, which cannot be explained by thermal expansion effects alone.\cite{Dounas-Frazer2013AJP} These comments often referred back to the exploratory activities that students completed during the first class, such as the station which involved observing and documenting the shapes of hanging chains and wires, both in loaded and unloaded conditions.
  
The focus of Classes~5--6 was on characterization of effects not included in students' original model. During Class~5, students discussed potential sources of uncertainty and systematic bias in their measurements. By the end of the class, students reached consensus that they would split into three groups to perform the following tests:
\begin{itemize}
\item modify the apparatus in order to facilitate more precise measurements of the height of the hanging mass, and repeat the experiment with the original procedure;
\item measure the elastic stretching of a room temperature wire under loads of various mass in order to determine whether elastic effects are negligible compared to thermal ones;
\item measure changes in height of the hanging mass at room temperature after repeated cycles of heating and cooling the wire, all in order to determine whether plastic deformation of the wire can be neglected.
\end{itemize}
These follow-up tests---which involve refinements of apparatus, model, and data-taking procedures---were conducted during Class~6. Collectively, they form the second iteration of the thermal expansion experiment. The first group modified the experimental apparatus by placing a small mirror behind the hanging mass in order to minimize parallax errors when using a ruler to measure the height of the load. The second group found that they must use a load that is sufficiently heavy that the wire hangs in a triangle (i.e., the mass of the wire is negligible in comparison), but sufficiently light that thermal expansion dominates elastic stretching for temperatures achievable in the classroom. The third group determined that the effects from the stretching out of kinks in the wire over time were negligible given the measurement uncertainties.

The results from Class~6 informed students' third and final iteration of the experiment, conducted during Class~7. Through a combination of small- and large-group discussions, the class decided to incorporate elastic stretching into their model for the length of the wire. In addition, they chose to repeat the experiment with a heavier mass. Thus, the third iteration of the experiment involved refinements of both the model and apparatus.

During their final discussion of the experiment, students discussed the iterative process of using data and uncertainty to inform improvements on their model and experimental procedure.  A key discussion topic was the role of iterative improvement as a general feature of scientific research.

\subsection{Final research projects}

In the second half of the course, students engaged in seven-week-long independent research projects (Fig.~\ref{fig:1}). The purpose of these projects was to provide students with additional opportunities to engage in the iterative process of experimental physics, in alignment with our first objective for the course. The details of research projects were highly idiosyncratic to the interests of the students involved, the resources available at Berkeley, and other context-dependent factors. Therefore, we only provide a brief description of the projects, focusing on aspects that could be implemented at other institutions.

During the eighth week of instruction, students were provided with a list of potential research topics. During Class~8, they self-sorted into groups of 3--4 students who shared a common interest in a particular topic. After students had chosen their groups and research topics, each group was paired with a graduate student volunteer who acted as a research advisor. Once per week during Classes~9--14, students met with their research advisors to discuss progress on their projects.

Under the guidance of their research advisors, students generated their own research projects related to their topic of interest. The research projects generated by students involved:
\begin{itemize}
\item measuring the thickness of oil on water using thin-film interference patterns;
\item measuring the speed of light using a microwave and marshmallows;
\item measuring the angular velocity of the earth's rotation about its axis using shadows; and,
\item comparing diffusion rates in one and two dimensions using scents in tubes and ink blots on paper.
\end{itemize}
An example of the apparatus developed to measure diffusion rates in two dimensions is shown in Fig.~\ref{fig:3}.

Each group designed a unique experiment that was iteratively improved through multiple initial trials. In many cases, early results forced students to revise their initial research goals. As the project deadline approached, students analyzed the data they had collected and compiled their conclusions into a final report. Students communicated their findings to the broader Compass community via a poster session at the end of the semester.


\subsection{Guided student reflections}

As part of their homework for {Intro to Measurement}, students were required to write weekly reflections about their college experiences. In other contexts, student reflection has been connected  to students' development of problem-solving skills,\cite{Mason2010} content knowledge,\cite{Scott2007}  conceptual understanding,\cite{May2002} and attitudes about physics.\cite{Ward2014} In particular, reflection has been recognized as beneficial for learning in laboratory contexts.~\cite{NASEM2015} The purpose of reflection in our course was to focus students on the iterative improvement of themselves as learners, by helping them develop and improve their lifelong learning skills.

Weekly reflections were assigned only during the first 11 weeks of the course so that students could focus on their final projects during the final 3 weeks  (Fig.~\ref{fig:1}). To help guide their reflections, students were given a rubric of 10 lifelong learning skills such as organization, collaboration, persistence, and self-compassion. The rubric was designed to encourage students to monitor their progress in developing these skills, set goals, and make plans to meet their goals---i.e., to support the types of iterative reflective practices associated with self-regulated learning.\cite{Zimmerman2002} To this end, the rubric defined each of the lifelong learning skills as well as what it means to be beginning, developing, or succeeding at the practice of each skill. The rubric further included prompt questions which students could use to help structure their reflections. The full rubric is included in the supplementary materials for this article.\footnote{\url{http://www.berkeleycompassproject.org/wordpress/wp-content/uploads/2012/06/GandhiEPAPS-rubric.pdf}}

In addition to weekly reflections, students were required to write a final, summative reflection at the end of the semester (Class 14). The final reflection included additional prompts to the questions listed on the rubric. In their final reflection, students were asked to: (1) describe their growth as learners over the course of the semester, drawing on grades and their weekly reflections as two complementary measures of growth; and (2) discuss sources of uncertainty and/or bias in these two measures of growth. One goal of the final reflections was to encourage students to reflect on their growth on timescales longer than one week. A second goals was for students to use experimental physics concepts like uncertainty and bias when evaluating their grades and reflections.

Timely, supportive feedback has been shown to promote self-regulated learning.\cite{Hattie2007} In addition, students' perception that they are being groomed for success (rather than weeded out) has been shown to promote retention in the sciences.\cite{Seymour1997} Accordingly, instructors responded to students' weekly reflections with individualized feedback that provided both instructive and affective support (final reflections did not receive feedback). Instructive feedback typically involved suggesting strategies and pointing students to resources. Consider the following example of strategic feedback:
\begin{quote}
``[M]ake sure you get the most out of every problem you do. What I mean is to take a few moments after you read a problem and think generally about what concepts or techniques might be useful before you start working. After you finish the problem, reflect on what you did."
\end{quote}
In this example, the instructor encouraged the use of metacognitive strategies like planning and reflecting.

Affective feedback focused on supporting students through personal struggles they faced. In these instances, instructors wrote responses that were affirming, empathizing, or normalizing in nature. For example:
\begin{quote}
``Figuring out that it is okay and even beneficial to drop a class is something that is hard to deal with. I know this is something I struggled with as a student."
\end{quote}
Here the instructor empathized with their student's difficult decision to drop a class. Further, the instructor normalized the student's experience by stating that the instructor, too, has had to come to terms with dropping classes. Almost all instructor responses included some aspect of affective and instructive support.

Below, we characterize the content of the weekly reflections. In addition, we highlight reflections from two students, Taylor and Micah, to illustrate how this activity facilitated iterative improvement of self.

\subsubsection{Content of weekly reflections}
To characterize the content of the student reflections, we analyzed 73 weekly reflections completed by 8 students over 11 weeks (completion rate of 83\% of 88 possible reflections). In order to characterize what skills students were reflecting on, we looked for explicit mention of rubric-related skills. When skills were not mentioned explicitly, we looked for rubric-related language from which a skill could be inferred. In 54 of 73 reflections, students explicitly stated the skills they were focused on (e.g., saying ``this week I will focus on self-compassion''). In 12 of the remaining 19 reflections, students used language directly from the rubrics without explicitly stating the skill (e.g., writing about ``frustration'' reflects language in the persistence skill prompt). The remaining 7 reflections did not follow the prompt for any particular skill. Students were not limited to one skill per reflection; in fact, almost a third of all reflections contained two or more learning skills.

The most popular skills for reflection were organization (35\% of 73 total reflections), collaboration (18\%), persistence (15\%), self-compassion (9\%), and courage (8\%). The other skills were only reflected on a few times each. Thus, more than half of the reflections focused on just three skills, namely, organization, collaboration, and persistence. These trends are consistent with other work that demonstrates that first-year college students struggle with organization generally and time management in particular.\cite{Reid2008,Clark2005} Indeed, reflections related to organization focused primarily on time management.

Most reflections about organization referred to time management specifically; others used phrases like, ``I want to be efficient with the little time I've got,'' or ``I did not have the time to keep up with my work.'' Most students were not used to the workload of college, and struggled to keep up. Students made statements such as:
\begin{quote}
``Due to the volume of work to be done (and the limited length of time), I sometimes find myself going through the motions \ldots"
\end{quote}
and,
\begin{quote}
``I stayed up all night and managed to get 2 and a half hours of sleep last night."
\end{quote}
Reflections about collaboration focused on working with other students, such as in a study group, or during office hours. In contrast, reflections about persistence did not have any clear themes. Students reflected on many different areas, such as: not giving up on long homework assignments, personal difficulties staying focused on schoolwork, and continuing towards a college degree despite feeling unsuccessful.

\subsubsection{Taylor's weekly reflections}
Taylor engaged in iterative improvement of self over the course of a multi-week interaction between with one of the instructors. Through this interaction, the student was able to shift their model of learning from seeing \emph{struggle as a sign of weakness} to seeing \emph{struggle as a sign that learning was taking place}. We provide two snapshots of Taylor's reflections, from Classes~3 and 6, to show the type of change that took place. We also provide one example of teacher feedback that seemed pivotal in supporting Taylor's iteration on their model of learning.

During their reflection for Class~3, Taylor described feeling self-doubt and uncertainty when they would get stuck on a physics problem:
\begin{quote}
``Self-doubting  still  haunts  me  whenever  I  am  trying  to  solve  physics problems. `Am I thinking this right?' this is a big question to me every time I am stuck in solving certain problems. It gives a strong sense of insecurity that my logic or intuition is wrong. I will convince myself to get help from others and get my problem resolved. But not quite, because I will end up thinking why I did not think of that or in the same way as they see the problems. Also, before I know that I answer a question correctly, uncertainty is still instilled in me. Another question hits me, `How can I know that I am correct?' If during exam I have no one to refer or to compare with, how can I know that I am right?" (Taylor, Class~3)
\end{quote}
The model of learning that seems to underlie this exposition is that success comes quickly for physicists (and capable physics students), and that one should feel certain in their answers. In the context of this model of learning, Taylor felt haunted by self-doubt when they got stuck solving a problem, which was associated with a feeling of insecurity. This self-doubt is also evident in the way Taylor spoke about seeking help from other students: rather than feeling better that the problem was resolved, Taylor questioned their own abilities (``I will end up thinking why I did not think of that").

Taylor and their instructor discussed these feelings over a number of weeks. For example, after Class~5, the instructor provided the following feedback:
\begin{quote}
``[I]t took me a while to understand when I was an undergrad was that the feeling of struggling with material is actually a good thing. Every time you struggle in a class, you learn something. It is important to struggle with problems and concepts to develop as a physicist. You also want to use your peers and instructors as resources to make the struggle as efficient and effective as possible. You don't have to do everything alone, although struggling for a short amount of time alone can be beneficial." (Instructor, Class~5)
\end{quote}
In this response the instructor \emph{empathized} with Taylor, saying he (the instructor) had similar feelings as an undergraduate. The instructor also worked to \emph{normalize} struggle, saying that struggle is an important part of learning and developing as a physicist. This response resonated with Taylor, and helped them iterate on their model of learning. This iteration is evident in Taylor's reflection from the following class, which Taylor wrote after taking a physics midterm:
\begin{quote}
``Last week midterm was tough, as expected. I struggled, like some of my classmates.
The moment I felt upset for underperformance, your words reminded me that struggle
is good; it means that youÕre learning something. Immediately I avoided my tendency
to get frustrated and tried to study my mistakes and think of a better thinking skill to tackle problems." (Taylor, Class~6)
\end{quote}
Taylor's response to struggling was now markedly different from before. Rather than taking struggle as a sign of inadequacy, Taylor leveraged their struggles as an opportunity to learn more. 

This example is a clear instance of how reflection helped students iterate on their models of themselves as learners. Initially, Taylor viewed struggle as a sign of weakness, which was detrimental to their success. Through reflection and instructor feedback, Taylor revised their model for learning, instead viewing struggle as a tool for learning. This type of iteration allowed Taylor to improve their practices associated with learning, ultimately supporting them as a lifelong learner.

\subsubsection{Micah's final reflection}
While Taylor used the weekly reflections to iterate on their model of learning, Micah used the final reflection to discuss failure and iteration in learning using language and ideas from experimental physics. In their final reflection, Micah noted that tests provide a ``composite score" of both knowledge and other factors:
\begin{quote}
``[T]ests, to me, are experiments where the independent variable is my knowledge and the dependent variable is my raw score. The test is a tool that I use to measure my knowledge but for me, being a disabled student, is also a source of error. There are a lot of uncertainties: the writing of the test, the amount of accommodations that I have \ldots and the availability of my textbooks being in an accessible format. These systematic biases lead to my tests being a composite score of how I am advocating for myself in terms of accommodations, my knowledge of the material, and the educational institutionÕs ability to accommodate me and other disabled students."
\end{quote}
Here Micah drew connections between tests and experiments by identifying tests as tools for measuring knowledge. Micah further identified a mapping between accessibility issues and systematic biases, drawing on language and concepts from experimental physics in the process (i.e., independent and dependent variables, error, uncertainty, and systematic bias). According to Micah, the outcome of a test is an imperfect measurement of knowledge because Micah's ability to advocate for accommodations and the university's ability to provide accommodations also contribute to the ``composite score."

The idea that tests provide a ``composite score" of both knowledge and accommodations is similar to the idea that the thermal expansion apparatus provides a `composite measurement' of both thermal and elastic effects. Using Micah's mapping, improving the thermal expansion experiment by controlling for elastic effects is similar to improving a testing experience by securing appropriate accommodations. Indeed, Micah described precisely this iterative process for improving the testing experience in Calculus~I. In their final reflection, Micah noted that they failed a Calculus~I midterm in the Fall semester, prior to enrolling in {Intro to Measurement}:
\begin{quote}
``I understood how to do calculus but I just didnÕt have the necessary tools to `do the experiment', [so to] speak. ... Finally, after being recommended multiple times by the DSP [Disabled Students' Program] staff to drop the class, I did. It felt like I was giving up. The experiment had failed. \ldots When November came around, I knew I needed to get my accommodations ready for [the Spring] semester. I emailed the DSP staff and the California Department of Rehabilitation and coordinated the accommodations that I needed for the next semester. I would take [Calculus~I] again and attempt to succeed. I would rerun my experiment correcting for error. During the remainder of the Fall semester, I focused on getting ahead for [Calculus~I] and coordinating my accommodations."
\end{quote}

After failing the midterm, Micah dropped Calculus~I. However, Micah felt that their low test grades were not a pure measurement of their knowledge, but a reflection of inadequate accommodations. After dropping the course, Micah took steps to ``rerun" an improved version of the ``experiment" by re-enrolling in Calculus~I during the Spring, after putting in place the necessary accommodations for their disability. Thus, Micah was able to use a model of grades as measurement to reinterpret the meaning of failure in their initial attempt at Calculus~I.

Micah described their experience in the Spring as follows:
\begin{quote}
``When it came to the spring semester, I established my accommodation base, although it was fairly flawed in many ways, but still better than none at all. [Micah described specific accommodations.] When I took my first midterm, I got a B. \ldots I took the accommodation lesson (using proper tools) \ldots in order to get a score that was more accurate to my understanding of calculus."
\end{quote}
Here, Micah described a cycle of iterative improvement. In preparation for the first calculus test in the Spring semester, Micah minimized accessibility-related ``systematic biases" by establishing a base of accommodations. Micah referred to this improvement as ``using proper tools." By using the proper tools for their experiment, Micah was able to earn a B on the test. Since Micah had minimized systematic biases contributing to their test score, they felt that this grade was a more accurate measure of their knowledge of caclulus.

In Micah's final reflection, we see how the ideas of iterative improvement of experiment and self can be intertwined to form self-narratives that allow for the reinterpretation of failure. Through viewing test scores as an approximation of their knowledge rather than a true representation, Micah constructed counter-narratives about their own ability to succeed as a physics student.

\section{Summary}
\label{sec:concl}
We have described the design and implementation of {Intro to Measurement}, an introductory laboratory course in which first-year undergraduate students engage in two types of iterative improvement: on experiments and on oneself. First, the course provided students multiple opportunities to engage in the type of iterative, model-based experimental physics tasks that are more typically found in upper-division lab courses\cite{Zwickl2014,Zwickl2013AJP} than in introductory ones.\cite{Wieman}

Second, {Intro to Measurement} provided students with opportunities to engage in iterative cycles of reflection, feedback, and growth---a process that is connected to self-regulated learning\cite{Zimmerman2002,Hattie2007} and retention in the sciences.\cite{Seymour1997} During weekly reflections, students wrote about their development of one or more lifelong learning skills, with a major focus on organization, collaboration, and persistence. We demonstrated how individualized feedback from instructors supported one student, Taylor, in making changes to their understanding of the relationship between struggling and learning.

Finally, we showed how a second student, Micah, synthesized the ideas of iterative improvement of experiment and self to construct a narrative about an experience of failure in an educational context. In this narrative, Micah interpreted test scores as composite measurements of both knowledge and issues related to accommodation of their disability. For this student, iterative improvement of self involved minimizing systematic bias in test scores by addressing their accommodation needs.

Thus, {Intro to Measurement} is an example of an introductory laboratory course that engaged students in authentic physics practices while simultaneously attending to the social and personal aspects of learning that are important for first-year undergraduate students. Nevertheless, several open questions remain. To what extent does this course format support all students in drawing  connections between experimental physics and learning, as Micah has done? And how might such connections benefit students, especially those from marginalized groups, in reinterpreting experiences of failure, advocating for institutional support, and persisting in the physics major? While these questions are beyond the scope of the present work, they may inform longitudinal studies of {Intro to Measurement} in the future.

\begin{acknowledgments} The authors acknowledge Andy diSessa for guidance throughout the project, Angela Little and Joel Corbo for feedback on this manuscript, and Jon Bender for creating an initial version of the rubric of lifelong learning skills. This work was supported by Compass, which in turn receives support from the University of California at Berkeley as well as from private donations. In addition, this work was supported by NSF grant DUE-1323101 and the AAU Undergraduate STEM Education Initiative.
\end{acknowledgments}

%


\begin{thebibliography}{36}%
\makeatletter
\providecommand \@ifxundefined [1]{%
 \@ifx{#1\undefined}
}%
\providecommand \@ifnum [1]{%
 \ifnum #1\expandafter \@firstoftwo
 \else \expandafter \@secondoftwo
 \fi
}%
\providecommand \@ifx [1]{%
 \ifx #1\expandafter \@firstoftwo
 \else \expandafter \@secondoftwo
 \fi
}%
\providecommand \natexlab [1]{#1}%
\providecommand \enquote  [1]{``#1''}%
\providecommand \bibnamefont  [1]{#1}%
\providecommand \bibfnamefont [1]{#1}%
\providecommand \citenamefont [1]{#1}%
\providecommand \href@noop [0]{\@secondoftwo}%
\providecommand \href [0]{\begingroup \@sanitize@url \@href}%
\providecommand \@href[1]{\@@startlink{#1}\@@href}%
\providecommand \@@href[1]{\endgroup#1\@@endlink}%
\providecommand \@sanitize@url [0]{\catcode `\\12\catcode `\$12\catcode
  `\&12\catcode `\#12\catcode `\^12\catcode `\_12\catcode `\%12\relax}%
\providecommand \@@startlink[1]{}%
\providecommand \@@endlink[0]{}%
\providecommand \url  [0]{\begingroup\@sanitize@url \@url }%
\providecommand \@url [1]{\endgroup\@href {#1}{\urlprefix }}%
\providecommand \urlprefix  [0]{URL }%
\providecommand \Eprint [0]{\href }%
\providecommand \doibase [0]{http://dx.doi.org/}%
\providecommand \selectlanguage [0]{\@gobble}%
\providecommand \bibinfo  [0]{\@secondoftwo}%
\providecommand \bibfield  [0]{\@secondoftwo}%
\providecommand \translation [1]{[#1]}%
\providecommand \BibitemOpen [0]{}%
\providecommand \bibitemStop [0]{}%
\providecommand \bibitemNoStop [0]{.\EOS\space}%
\providecommand \EOS [0]{\spacefactor3000\relax}%
\providecommand \BibitemShut  [1]{\csname bibitem#1\endcsname}%
\let\auto@bib@innerbib\@empty
\bibitem [{\citenamefont {Brewe}(2008)}]{Brewe2008}%
  \BibitemOpen
  \bibfield  {author} {\bibinfo {author} {\bibfnamefont {Eric}\ \bibnamefont
  {Brewe}},\ }\bibfield  {title} {\enquote {\bibinfo {title} {Modeling theory
  applied: Modeling instruction in introductory physics},}\ }\href {\doibase
  http://dx.doi.org/10.1119/1.2983148} {\bibfield  {journal} {\bibinfo
  {journal} {Am. J. of Phys.}\ }\textbf {\bibinfo {volume} {76}},\ \bibinfo
  {pages} {1155--1160} (\bibinfo {year} {2008})}\BibitemShut {NoStop}%
\bibitem [{\citenamefont {Etkina}\ and\ \citenamefont {van
  Heuvelen}(2007)}]{Etkina2007}%
  \BibitemOpen
  \bibfield  {author} {\bibinfo {author} {\bibfnamefont {Eugenia}\ \bibnamefont
  {Etkina}}\ and\ \bibinfo {author} {\bibfnamefont {Alan}\ \bibnamefont {van
  Heuvelen}},\ }\bibfield  {title} {\enquote {\bibinfo {title} {Investigative
  science learning environment - a science process approach to learning
  physics},}\ }in\ \href@noop {} {\emph {\bibinfo {booktitle} {Research-Based
  Reform of University Physics}}},\ Vol.~\bibinfo {volume} {1}\ (\bibinfo
  {year} {2007})\BibitemShut {NoStop}%
\bibitem [{\citenamefont {Zwickl}\ \emph {et~al.}(2014)\citenamefont {Zwickl},
  \citenamefont {Finkelstein},\ and\ \citenamefont {Lewandowski}}]{Zwickl2014}%
  \BibitemOpen
  \bibfield  {author} {\bibinfo {author} {\bibfnamefont {Benjamin~M.}\
  \bibnamefont {Zwickl}}, \bibinfo {author} {\bibfnamefont {Noah}\ \bibnamefont
  {Finkelstein}}, \ and\ \bibinfo {author} {\bibfnamefont {H.~J.}\ \bibnamefont
  {Lewandowski}},\ }\bibfield  {title} {\enquote {\bibinfo {title}
  {Incorporating learning goals about modeling into an upper-division physics
  laboratory experiment},}\ }\href {\doibase
  http://dx.doi.org/10.1119/1.4875924} {\bibfield  {journal} {\bibinfo
  {journal} {Am. J. of Phys.}\ }\textbf {\bibinfo {volume} {82}},\ \bibinfo
  {pages} {876--882} (\bibinfo {year} {2014})}\BibitemShut {NoStop}%
\bibitem [{\citenamefont {Zwickl}\ \emph {et~al.}(2013)\citenamefont {Zwickl},
  \citenamefont {Finkelstein},\ and\ \citenamefont
  {Lewandowski}}]{Zwickl2013AJP}%
  \BibitemOpen
  \bibfield  {author} {\bibinfo {author} {\bibfnamefont {Benjamin~M.}\
  \bibnamefont {Zwickl}}, \bibinfo {author} {\bibfnamefont {Noah}\ \bibnamefont
  {Finkelstein}}, \ and\ \bibinfo {author} {\bibfnamefont {Heather~J.}\
  \bibnamefont {Lewandowski}},\ }\bibfield  {title} {\enquote {\bibinfo {title}
  {The process of transforming an advanced lab course: Goals, curriculum, and
  assessments},}\ }\href {\doibase http://dx.doi.org/10.1119/1.4768890}
  {\bibfield  {journal} {\bibinfo  {journal} {Am. J. of Phys.}\ }\textbf
  {\bibinfo {volume} {81}},\ \bibinfo {pages} {63--70} (\bibinfo {year}
  {2013})}\BibitemShut {NoStop}%
\bibitem [{\citenamefont {{AAPT Committee on Laboratories}}(2015)}]{AAPT2015}%
  \BibitemOpen
  \bibfield  {author} {\bibinfo {author} {\bibnamefont {{AAPT Committee on
  Laboratories}}},\ }\href@noop {} {\enquote {\bibinfo {title} {{AAPT
  Recommendations for the Undergraduate Physics Laboratory Curriculum}},}\ }
  (\bibinfo {year} {2015})\BibitemShut {NoStop}%
\bibitem [{\citenamefont {Wieman}(2015)}]{Wieman}%
  \BibitemOpen
  \bibfield  {author} {\bibinfo {author} {\bibfnamefont {Carl}\ \bibnamefont
  {Wieman}},\ }\bibfield  {title} {\enquote {\bibinfo {title} {Comparative
  cognitive task analyses of experimental science and instructional laboratory
  courses},}\ }\href {\doibase 10.1119/1.4928349} {\bibfield  {journal}
  {\bibinfo  {journal} {The Physics Teacher}\ }\textbf {\bibinfo {volume}
  {53}},\ \bibinfo {pages} {349--351} (\bibinfo {year} {2015})}\BibitemShut
  {NoStop}%
\bibitem [{\citenamefont {{NGSS Lead States}}(2013)}]{NGSS2013}%
  \BibitemOpen
  \bibfield  {author} {\bibinfo {author} {\bibnamefont {{NGSS Lead States}}},\
  }\href@noop {} {\enquote {\bibinfo {title} {{Next Generation Science
  Standards: For States, By States}},}\ } (\bibinfo {year} {2013})\BibitemShut
  {NoStop}%
\bibitem [{\citenamefont {Reid}\ and\ \citenamefont {Moore}(2008)}]{Reid2008}%
  \BibitemOpen
  \bibfield  {author} {\bibinfo {author} {\bibfnamefont {M.~Jeanne}\
  \bibnamefont {Reid}}\ and\ \bibinfo {author} {\bibfnamefont {James~L.}\
  \bibnamefont {Moore}},\ }\bibfield  {title} {\enquote {\bibinfo {title}
  {College readiness and academic preparation for postsecondary education: Oral
  histories of first-generation urban college students},}\ }\href {\doibase
  10.1177/0042085907312346} {\bibfield  {journal} {\bibinfo  {journal} {Urban
  Education}\ }\textbf {\bibinfo {volume} {43}},\ \bibinfo {pages} {240--261}
  (\bibinfo {year} {2008})}\BibitemShut {NoStop}%
\bibitem [{\citenamefont {Dounas-Frazer}\ and\ \citenamefont
  {Reinholz}(2015)}]{Dounas-Frazer2015}%
  \BibitemOpen
  \bibfield  {author} {\bibinfo {author} {\bibfnamefont {Dimitri~R.}\
  \bibnamefont {Dounas-Frazer}}\ and\ \bibinfo {author} {\bibfnamefont
  {Daniel~L.}\ \bibnamefont {Reinholz}},\ }\bibfield  {title} {\enquote
  {\bibinfo {title} {Attending to lifelong learning skills through guided
  reflection in a physics class},}\ }\href {\doibase
  http://dx.doi.org/10.1119/1.4930083} {\bibfield  {journal} {\bibinfo
  {journal} {American Journal of Physics}\ }\textbf {\bibinfo {volume} {83}},\
  \bibinfo {pages} {881--891} (\bibinfo {year} {2015})}\BibitemShut {NoStop}%
\bibitem [{\citenamefont {Jaber}(2014)}]{Jaber2014}%
  \BibitemOpen
  \bibfield  {author} {\bibinfo {author} {\bibfnamefont {Lama~Ziad}\
  \bibnamefont {Jaber}},\ }\emph {\bibinfo {title} {{Affective dynamics of
  students' disciplinary engagement in science}}},\ \href@noop {} {\bibinfo
  {type} {{Ph.D.} thesis}},\ \bibinfo  {school} {Tufts University} (\bibinfo
  {year} {2014})\BibitemShut {NoStop}%
\bibitem [{\citenamefont {on~Undergraduate Physics Education~Research}\ \emph
  {et~al.}(2013)\citenamefont {on~Undergraduate Physics Education~Research},
  \citenamefont {on~Physics}, \citenamefont {on~Engineering},\ and\
  \citenamefont {Council}}]{NRC2013}%
  \BibitemOpen
  \bibfield  {author} {\bibinfo {author} {\bibfnamefont {Committee}\
  \bibnamefont {on~Undergraduate Physics Education~Research}}, \bibinfo
  {author} {\bibfnamefont {Implementation;~Board}\ \bibnamefont {on~Physics}},
  \bibinfo {author} {\bibfnamefont {Astronomy;~Division}\ \bibnamefont
  {on~Engineering}}, \ and\ \bibinfo {author} {\bibfnamefont {Physical
  Sciences; National~Research}\ \bibnamefont {Council}},\ }\href
  {http://www.nap.edu/openbook.php?record_id=18312} {\emph {\bibinfo {title}
  {Adapting to a Changing World--Challenges and Opportunities in Undergraduate
  Physics Education}}}\ (\bibinfo  {publisher} {The National Academies Press},\
  \bibinfo {year} {2013})\BibitemShut {NoStop}%
\bibitem [{\citenamefont {{National Academy of Sciences, National Academy of
  Engineering and Institute of Medicine Committee on Underrepresented Groups
  and the Expansion of the Science and Engineering Workforce
  Pipeline}}(2011)}]{NCES2011}%
  \BibitemOpen
  \bibfield  {author} {\bibinfo {author} {\bibnamefont {{National Academy of
  Sciences, National Academy of Engineering and Institute of Medicine Committee
  on Underrepresented Groups and the Expansion of the Science and Engineering
  Workforce Pipeline}}},\ }\href {http://www.ncbi.nlm.nih.gov/books/NBK83369/}
  {\emph {\bibinfo {title} {Expanding Underrepresented Minority
  Participation}}}\ (\bibinfo  {publisher} {National Academies Press},\
  \bibinfo {year} {2011})\ Chap.\ \bibinfo {chapter} {6, Academic and Social
  Support}\BibitemShut {NoStop}%
\bibitem [{\citenamefont {Clark}(2005)}]{Clark2005}%
  \BibitemOpen
  \bibfield  {author} {\bibinfo {author} {\bibfnamefont {Marcia~Roe}\
  \bibnamefont {Clark}},\ }\bibfield  {title} {\enquote {\bibinfo {title}
  {Negotiating the freshman year: Challenges and strategies among first-year
  college students},}\ }\href {\doibase 10.1353/csd.2005.0022} {\bibfield
  {journal} {\bibinfo  {journal} {Journal of College Student Development}\
  }\textbf {\bibinfo {volume} {46}},\ \bibinfo {pages} {296--316} (\bibinfo
  {year} {2005})}\BibitemShut {NoStop}%
\bibitem [{\citenamefont {Zwickl}\ \emph {et~al.}(2015)\citenamefont {Zwickl},
  \citenamefont {Hu}, \citenamefont {Finkelstein},\ and\ \citenamefont
  {Lewandowski}}]{Zwickl2015}%
  \BibitemOpen
  \bibfield  {author} {\bibinfo {author} {\bibfnamefont {Benjamin~M.}\
  \bibnamefont {Zwickl}}, \bibinfo {author} {\bibfnamefont {Dehui}\
  \bibnamefont {Hu}}, \bibinfo {author} {\bibfnamefont {Noah}\ \bibnamefont
  {Finkelstein}}, \ and\ \bibinfo {author} {\bibfnamefont {H.~J.}\ \bibnamefont
  {Lewandowski}},\ }\bibfield  {title} {\enquote {\bibinfo {title} {Model-based
  reasoning in the physics laboratory: Framework and initial results},}\ }\href
  {\doibase 10.1103/PhysRevSTPER.11.020113} {\bibfield  {journal} {\bibinfo
  {journal} {Phys. Rev. ST Phys. Educ. Res.}\ }\textbf {\bibinfo {volume}
  {11}},\ \bibinfo {pages} {020113} (\bibinfo {year} {2015})}\BibitemShut
  {NoStop}%
\bibitem [{\citenamefont {Dounas-Frazer}\ \emph {et~al.}(in press)\citenamefont
  {Dounas-Frazer}, \citenamefont {Bogart}, \citenamefont {Stetzer},\ and\
  \citenamefont {Lewandowski}}]{Dounas-Frazer2015arXiv}%
  \BibitemOpen
  \bibfield  {author} {\bibinfo {author} {\bibfnamefont {Dimitri~R.}\
  \bibnamefont {Dounas-Frazer}}, \bibinfo {author} {\bibfnamefont {Kevin L.
  Van~De}\ \bibnamefont {Bogart}}, \bibinfo {author} {\bibfnamefont
  {MacKenzie~R.}\ \bibnamefont {Stetzer}}, \ and\ \bibinfo {author}
  {\bibfnamefont {H.~J.}\ \bibnamefont {Lewandowski}},\ }\bibfield  {title}
  {\enquote {\bibinfo {title} {The role of modeling in troubleshooting: an
  example from electronics},}\ }in\ \href@noop {} {\emph {\bibinfo {booktitle}
  {Physics Education Research Conference 2015}}},\ \bibinfo {series and number}
  {PER Conference}\ (\bibinfo {address} {Minneapolis, MN},\ \bibinfo {year} {in
  press})\BibitemShut {NoStop}%
\bibitem [{\citenamefont {Etkina}\ \emph {et~al.}(2008)\citenamefont {Etkina},
  \citenamefont {Karelina},\ and\ \citenamefont
  {Ruibal-Villasenor}}]{Etkina2008}%
  \BibitemOpen
  \bibfield  {author} {\bibinfo {author} {\bibfnamefont {Eugenia}\ \bibnamefont
  {Etkina}}, \bibinfo {author} {\bibfnamefont {Anna}\ \bibnamefont {Karelina}},
  \ and\ \bibinfo {author} {\bibfnamefont {Maria}\ \bibnamefont
  {Ruibal-Villasenor}},\ }\bibfield  {title} {\enquote {\bibinfo {title} {How
  long does it take? a study of student acquisition of scientific abilities},}\
  }\href {\doibase 10.1103/PhysRevSTPER.4.020108} {\bibfield  {journal}
  {\bibinfo  {journal} {Phys. Rev. ST Phys. Educ. Res.}\ }\textbf {\bibinfo
  {volume} {4}},\ \bibinfo {pages} {020108} (\bibinfo {year}
  {2008})}\BibitemShut {NoStop}%
\bibitem [{\citenamefont {Etkina}\ \emph {et~al.}(2009)\citenamefont {Etkina},
  \citenamefont {Karelina}, \citenamefont {Murthy},\ and\ \citenamefont
  {Ruibal-Villasenor}}]{Etkina2009}%
  \BibitemOpen
  \bibfield  {author} {\bibinfo {author} {\bibfnamefont {Eugenia}\ \bibnamefont
  {Etkina}}, \bibinfo {author} {\bibfnamefont {Anna}\ \bibnamefont {Karelina}},
  \bibinfo {author} {\bibfnamefont {Sahana}\ \bibnamefont {Murthy}}, \ and\
  \bibinfo {author} {\bibfnamefont {Maria}\ \bibnamefont {Ruibal-Villasenor}},\
  }\bibfield  {title} {\enquote {\bibinfo {title} {Using action research to
  improve learning and formative assessment to conduct research},}\ }\href
  {\doibase 10.1103/PhysRevSTPER.5.010109} {\bibfield  {journal} {\bibinfo
  {journal} {Phys. Rev. ST Phys. Educ. Res.}\ }\textbf {\bibinfo {volume}
  {5}},\ \bibinfo {pages} {010109} (\bibinfo {year} {2009})}\BibitemShut
  {NoStop}%
\bibitem [{\citenamefont {Jackson}(1968)}]{Jackson1968}%
  \BibitemOpen
  \bibfield  {author} {\bibinfo {author} {\bibfnamefont {P.~W.}\ \bibnamefont
  {Jackson}},\ }\href@noop {} {\emph {\bibinfo {title} {Life in {Classrooms}.
  {New} {York}: {Holt}, {Rinehart} and {Winston}}}}\ (\bibinfo  {publisher}
  {Holt, Rinehart and Winston},\ \bibinfo {address} {New York},\ \bibinfo
  {year} {1968})\BibitemShut {NoStop}%
\bibitem [{Note1()}]{Note1}%
  \BibitemOpen
  \bibinfo {note} {One of the authors (DRDF) co-designed and co-taught the
  initial iteration of {Intro to Measurement} in the spring semester of 2012,
  and two of the authors (PRG and JAL) refined and co-taught the course the
  following spring. We report on the implementation and pilot study of the
  second iteration of the course.}\BibitemShut {Stop}%
\bibitem [{\citenamefont {Albanna}\ \emph {et~al.}(2013)\citenamefont
  {Albanna}, \citenamefont {Corbo}, \citenamefont {Dounas-Frazer},
  \citenamefont {Little},\ and\ \citenamefont {Zaniewski}}]{Albanna2013}%
  \BibitemOpen
  \bibfield  {author} {\bibinfo {author} {\bibfnamefont {Badr~F.}\ \bibnamefont
  {Albanna}}, \bibinfo {author} {\bibfnamefont {Joel~C.}\ \bibnamefont
  {Corbo}}, \bibinfo {author} {\bibfnamefont {Dimitri~R.}\ \bibnamefont
  {Dounas-Frazer}}, \bibinfo {author} {\bibfnamefont {Angela}\ \bibnamefont
  {Little}}, \ and\ \bibinfo {author} {\bibfnamefont {Anna~M.}\ \bibnamefont
  {Zaniewski}},\ }\bibfield  {title} {\enquote {\bibinfo {title} {Building
  classroom and organizational structure around positive cultural values},}\
  }\href {\doibase 10.1063/1.4789638} {\bibfield  {journal} {\bibinfo
  {journal} {AIP Conference Proceedings}\ }\textbf {\bibinfo {volume} {1513}},\
  \bibinfo {pages} {7--10} (\bibinfo {year} {2013})}\BibitemShut {NoStop}%
\bibitem [{\citenamefont {Dounas-Frazer}\ \emph
  {et~al.}(2013{\natexlab{a}})\citenamefont {Dounas-Frazer}, \citenamefont
  {Lynn}, \citenamefont {Zaniewski},\ and\ \citenamefont
  {Roth}}]{Dounas-Frazer2013TPT}%
  \BibitemOpen
  \bibfield  {author} {\bibinfo {author} {\bibfnamefont {D.~R.}\ \bibnamefont
  {Dounas-Frazer}}, \bibinfo {author} {\bibfnamefont {J.}~\bibnamefont {Lynn}},
  \bibinfo {author} {\bibfnamefont {A.~M.}\ \bibnamefont {Zaniewski}}, \ and\
  \bibinfo {author} {\bibfnamefont {N.}~\bibnamefont {Roth}},\ }\bibfield
  {title} {\enquote {\bibinfo {title} {Learning about non-newtonian fluids in a
  student-driven classroom},}\ }\href {\doibase 10.1119/1.4772035} {\bibfield
  {journal} {\bibinfo  {journal} {The Physics Teacher}\ }\textbf {\bibinfo
  {volume} {51}},\ \bibinfo {pages} {32--34} (\bibinfo {year}
  {2013}{\natexlab{a}})}\BibitemShut {NoStop}%
\bibitem [{\citenamefont {{American Institute of Physics}}(2015)}]{AIP2012}%
  \BibitemOpen
  \bibfield  {author} {\bibinfo {author} {\bibnamefont {{American Institute of
  Physics}}},\ }\bibfield  {title} {\enquote {\bibinfo {title} {{Physics
  Bachelor's Degrees: Results from the 2010 Survey of Enrollments and
  Degrees}},}\ }\href
  {http://www.aip.org/statistics/reports/physics-bachelors-degrees} {\bibfield
   {journal} {\bibinfo  {journal} {Focus On}\ } (\bibinfo {year}
  {2015})}\BibitemShut {NoStop}%
\bibitem [{Note2()}]{Note2}%
  \BibitemOpen
  \bibinfo {note} {Consent was obtained pursuant to protocol \#2012-11-4836,
  approved by the UC Berkeley Committee for Protection of Human
  Subjects.}\BibitemShut {Stop}%
\bibitem [{\citenamefont {Cohen}\ and\ \citenamefont
  {Lotan}(1997)}]{Cohen1997}%
  \BibitemOpen
  \bibinfo {editor} {\bibfnamefont {Elizabeth~G.}\ \bibnamefont {Cohen}}\ and\
  \bibinfo {editor} {\bibfnamefont {Rachel~A.}\ \bibnamefont {Lotan}},\ eds.,\
  \href@noop {} {\emph {\bibinfo {title} {{Working for Equity in Heterogeneous
  Classrooms: Sociological Theory in Practice}}}}\ (\bibinfo  {publisher}
  {Teachers College Press},\ \bibinfo {year} {1997})\BibitemShut {NoStop}%
\bibitem [{\citenamefont {Papert}\ and\ \citenamefont
  {Hardel}(1991)}]{Papert1991}%
  \BibitemOpen
  \bibfield  {author} {\bibinfo {author} {\bibfnamefont {Seymour}\ \bibnamefont
  {Papert}}\ and\ \bibinfo {author} {\bibfnamefont {Idit}\ \bibnamefont
  {Hardel}},\ }\href@noop {} {\emph {\bibinfo {title} {{Constructionism}}}}\
  (\bibinfo  {publisher} {Ablex Publishing},\ \bibinfo {year}
  {1991})\BibitemShut {NoStop}%
\bibitem [{\citenamefont {Planin\u{s}i\u{c}}(2007)}]{Planinsic2007}%
  \BibitemOpen
  \bibfield  {author} {\bibinfo {author} {\bibfnamefont {Gorazd}\ \bibnamefont
  {Planin\u{s}i\u{c}}},\ }\bibfield  {title} {\enquote {\bibinfo {title}
  {Project laboratory for first-year students},}\ }\href
  {http://stacks.iop.org/0143-0807/28/i=3/a=S07} {\bibfield  {journal}
  {\bibinfo  {journal} {European Journal of Physics}\ }\textbf {\bibinfo
  {volume} {28}},\ \bibinfo {pages} {S71} (\bibinfo {year} {2007})}\BibitemShut
  {NoStop}%
\bibitem [{\citenamefont {Dounas-Frazer}\ \emph
  {et~al.}(2013{\natexlab{b}})\citenamefont {Dounas-Frazer}, \citenamefont
  {Gandhi},\ and\ \citenamefont {Iwata}}]{Dounas-Frazer2013AJP}%
  \BibitemOpen
  \bibfield  {author} {\bibinfo {author} {\bibfnamefont {Dimitri~R.}\
  \bibnamefont {Dounas-Frazer}}, \bibinfo {author} {\bibfnamefont {Punit~R.}\
  \bibnamefont {Gandhi}}, \ and\ \bibinfo {author} {\bibfnamefont
  {Geoffrey~Z.}\ \bibnamefont {Iwata}},\ }\bibfield  {title} {\enquote
  {\bibinfo {title} {Uncertainty analysis for a simple thermal expansion
  experiment},}\ }\href {\doibase 10.1119/1.4789875} {\bibfield  {journal}
  {\bibinfo  {journal} {American Journal of Physics}\ }\textbf {\bibinfo
  {volume} {81}},\ \bibinfo {pages} {338--342} (\bibinfo {year}
  {2013}{\natexlab{b}})}\BibitemShut {NoStop}%
\bibitem [{\citenamefont {Mason}\ and\ \citenamefont
  {Singh}(2010)}]{Mason2010}%
  \BibitemOpen
  \bibfield  {author} {\bibinfo {author} {\bibfnamefont {Andrew}\ \bibnamefont
  {Mason}}\ and\ \bibinfo {author} {\bibfnamefont {Chandralekha}\ \bibnamefont
  {Singh}},\ }\bibfield  {title} {\enquote {\bibinfo {title} {Using reflection
  with peers to help students learn effective problem solving strategies},}\
  }\href {\doibase http://dx.doi.org/10.1063/1.3515243} {\bibfield  {journal}
  {\bibinfo  {journal} {AIP Conference Proceedings}\ }\textbf {\bibinfo
  {volume} {1289}},\ \bibinfo {pages} {41--44} (\bibinfo {year}
  {2010})}\BibitemShut {NoStop}%
\bibitem [{\citenamefont {Scott}\ \emph {et~al.}(2007)\citenamefont {Scott},
  \citenamefont {Stelzer},\ and\ \citenamefont {Gladding}}]{Scott2007}%
  \BibitemOpen
  \bibfield  {author} {\bibinfo {author} {\bibfnamefont {Michael~L.}\
  \bibnamefont {Scott}}, \bibinfo {author} {\bibfnamefont {Tim}\ \bibnamefont
  {Stelzer}}, \ and\ \bibinfo {author} {\bibfnamefont {Gary}\ \bibnamefont
  {Gladding}},\ }\bibfield  {title} {\enquote {\bibinfo {title} {Explicit
  reflection in an introductory physics course},}\ }\href {\doibase
  http://dx.doi.org/10.1063/1.2820929} {\bibfield  {journal} {\bibinfo
  {journal} {AIP Conference Proceedings}\ }\textbf {\bibinfo {volume} {951}},\
  \bibinfo {pages} {188--191} (\bibinfo {year} {2007})}\BibitemShut {NoStop}%
\bibitem [{\citenamefont {May}\ and\ \citenamefont {Etkina}(2002)}]{May2002}%
  \BibitemOpen
  \bibfield  {author} {\bibinfo {author} {\bibfnamefont {David~B.}\
  \bibnamefont {May}}\ and\ \bibinfo {author} {\bibfnamefont {Eugenia}\
  \bibnamefont {Etkina}},\ }\bibfield  {title} {\enquote {\bibinfo {title}
  {College physics students’ epistemological self-reflection and its
  relationship to conceptual learning},}\ }\href {\doibase
  http://dx.doi.org/10.1119/1.1503377} {\bibfield  {journal} {\bibinfo
  {journal} {American Journal of Physics}\ }\textbf {\bibinfo {volume} {70}},\
  \bibinfo {pages} {1249--1258} (\bibinfo {year} {2002})}\BibitemShut {NoStop}%
\bibitem [{\citenamefont {Ward}\ and\ \citenamefont {Duda}(2014)}]{Ward2014}%
  \BibitemOpen
  \bibfield  {author} {\bibinfo {author} {\bibfnamefont {K.}~\bibnamefont
  {Ward}}\ and\ \bibinfo {author} {\bibfnamefont {G.}~\bibnamefont {Duda}},\
  }\href
  {http://www.compadre.org/PER/perc/2014/files/Ward_Duda_PERC2014_reflection_ver3.pdf}
  {\emph {\bibinfo {title} {The role of student reflection in project-based
  learning physics courses}}}\ (\bibinfo  {publisher} {Proceedings of the 2014
  Physics Education Research Conference},\ \bibinfo {address} {Minneapolis,
  MN},\ \bibinfo {year} {2014})\BibitemShut {NoStop}%
\bibitem [{\citenamefont {{National Academies of Sciences, Engineering, and
  Medicine}}(2015)}]{NASEM2015}%
  \BibitemOpen
  \bibfield  {author} {\bibinfo {author} {\bibnamefont {{National Academies of
  Sciences, Engineering, and Medicine}}},\ }\href
  {http://www.nap.edu/catalog/21851/integrating-discovery-based-research-into-the-undergraduate-curriculum-report-of}
  {\emph {\bibinfo {title} {Integrating Discovery-Based Research into the
  Undergraduate Curriculum: Report of a Convocation}}}\ (\bibinfo  {publisher}
  {National Academies Press},\ \bibinfo {year} {2015})\ Chap.\ \bibinfo
  {chapter} {{3, Promising Practices and Ongoing Challenges}}\BibitemShut
  {NoStop}%
\bibitem [{\citenamefont {Zimmerman}(2002)}]{Zimmerman2002}%
  \BibitemOpen
  \bibfield  {author} {\bibinfo {author} {\bibfnamefont {Barry~J.}\
  \bibnamefont {Zimmerman}},\ }\bibfield  {title} {\enquote {\bibinfo {title}
  {Becoming a self-regulated learner: An overview},}\ }\href {\doibase
  10.1207/s15430421tip4102_2} {\bibfield  {journal} {\bibinfo  {journal}
  {Theory Into Practice}\ }\textbf {\bibinfo {volume} {41}},\ \bibinfo {pages}
  {64--70} (\bibinfo {year} {2002})}\BibitemShut {NoStop}%
\bibitem [{Note3()}]{Note3}%
  \BibitemOpen
  \bibinfo {note} {\protect \url
  {[URL will be inserted by AIP]}}\BibitemShut
  {NoStop}%
\bibitem [{\citenamefont {Hattie}\ and\ \citenamefont
  {Timperley}(2007)}]{Hattie2007}%
  \BibitemOpen
  \bibfield  {author} {\bibinfo {author} {\bibfnamefont {John}\ \bibnamefont
  {Hattie}}\ and\ \bibinfo {author} {\bibfnamefont {Helen}\ \bibnamefont
  {Timperley}},\ }\bibfield  {title} {\enquote {\bibinfo {title} {The power of
  feedback},}\ }\href {\doibase 10.3102/003465430298487} {\bibfield  {journal}
  {\bibinfo  {journal} {Review of Educational Research}\ }\textbf {\bibinfo
  {volume} {77}},\ \bibinfo {pages} {81--112} (\bibinfo {year}
  {2007})}\BibitemShut {NoStop}%
\bibitem [{\citenamefont {Seymour}\ and\ \citenamefont
  {Hewitt}(1997)}]{Seymour1997}%
  \BibitemOpen
  \bibfield  {author} {\bibinfo {author} {\bibfnamefont {Elaine}\ \bibnamefont
  {Seymour}}\ and\ \bibinfo {author} {\bibfnamefont {Nancy~M.}\ \bibnamefont
  {Hewitt}},\ }\href@noop {} {\emph {\bibinfo {title} {{Talking about leaving:
  Why undergraduates leave the sciences}}}}\ (\bibinfo  {publisher} {Westview
  Press},\ \bibinfo {year} {1997})\BibitemShut {NoStop}%
\end{thebibliography}

\begin{figure*}[t]\center
\caption{\label{fig:1} Timeline of course activities.}
\end{figure*}

\begin{figure}[t]\center
\caption{\label{fig:2} Thermal expansion apparatus. The ends of a wire are attached to blocks of wood via small metal hooks. Both blocks are fastened to a level, rigid surface (not shown) via c-clamps. Insulated copper wires are attached to the hooks to facilitate electrical connections to an ac power supply. Temperature is controlled by varying the current in the wire. This apparatus has been described in greater detail elsewhere.\cite{Dounas-Frazer2013AJP}}
\end{figure}

\begin{figure}[t]\center
\caption{\label{fig:3} Apparatus for measuring diffusion rates two dimensions using ink blots on paper. 
Students placed a drop of dye on a horizontal sheet of paper towel. They recorded the diffusion of the dye through the paper towel with a video camera and used their data to measure the spot size as a
function of time.}
\end{figure}

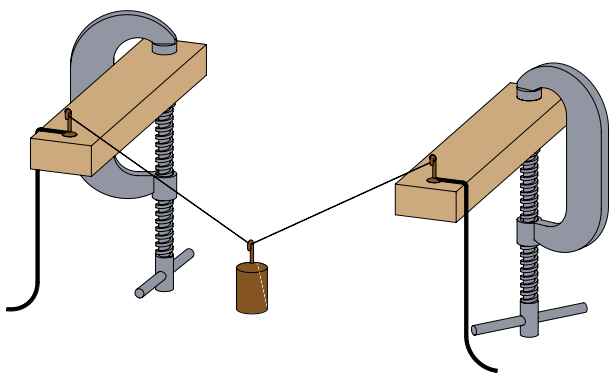
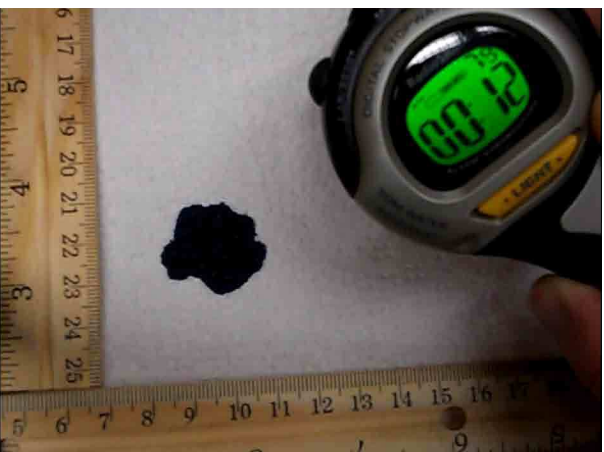
\end{document}